\pdfoutput=1
\documentclass[12pt]{article}
\usepackage{epsf,amsmath,amssymb,graphicx,dcolumn}
\usepackage{scalefnt,cite}
\usepackage{hyperref}
\usepackage{cleveref,etoolbox,color,soul}
\usepackage{listings,booktabs}
\usepackage[utf8]{inputenc}

\newcommand{\citere}[1]{Ref.\,\cite{#1}}
\newcommand{\citeres}[1]{Refs.\,\cite{#1}}
\newcommand{\code}{\tt}
\newcommand{\sushi}[1]{{\code SusHi #1}}
\newcommand{\abbrev}{\scalefont{.9}}
\newcommand{\eqn}[1]{Eq.\,(\ref{#1})}
\newcommand{\fig}[1]{Fig.\,\ref{#1}}
\newcommand{\tab}[1]{Tab.\,\ref{#1}}
\newcommand{\sct}[1]{Section~\ref{#1}}

\newcommand{\lhc}{{\abbrev LHC}}
\newcommand{\qcd}{{\abbrev QCD}}
\newcommand{\sm}{{\abbrev SM}}
\newcommand{\thdm}{{\abbrev 2HDM}}
\newcommand{\mssm}{{\abbrev MSSM}}
\newcommand{\atlas}{{\abbrev ATLAS}}
\newcommand{\cms}{{\abbrev CMS}}

\newcommand{\cp}{{\abbrev $\mathcal{CP}$}}
\newcommand{\lo}{{\abbrev LO}}
\newcommand{\nlo}{{\abbrev NLO}}
\newcommand{\nnlo}{{\abbrev NNLO}}

\begin{document}
\begin{flushleft}
\end{flushleft}
\vspace*{-1.35cm}
\begin{flushright}
{\tt DESY 17-173},
{\tt KA-TP-34-2017 }
\end{flushright}

\long\def\symbolfootnote[#1]#2{\begingroup%
\def\thefootnote{\fnsymbol{footnote}}\footnote[#1]{#2}\endgroup}

\vspace{0.1cm}

\begin{center}
\Large\bf\boldmath
$pp\to A\to Zh$ and the wrong-sign limit\\ of the Two-Higgs-Doublet Model
\unboldmath
\end{center}
\vspace{0.05cm}
\begin{center}
Pedro M. Ferreira$^{a,b}$,
Stefan Liebler$^c$,
Jonas Wittbrodt$^d$\symbolfootnote[0]{Electronic addresses:
pmmferreira@fc.ul.pt, stefan.liebler@kit.edu, jonas.wittbrodt@desy.de.}\\[0.4cm]
{\small
{\sl${}^a$Instituto Superior de Engenharia de Lisboa - ISEL, 1959-007 Lisboa,  Portugal}\\[0.2em]
{\sl${}^b$Centro  de  F{\'i}sica  Te{\'o}rica  e  Computacional,  Universidade  de  Lisboa, 1649-003 Lisboa, Portugal}\\[0.2em]
{\sl${}^c$Institute for Theoretical Physics, Karlsruhe Institute of Technology,
76131 Karlsruhe, Germany}\\[0.2em]
{\sl${}^d$DESY, Notkestra{\ss}e 85, 22607 Hamburg, Germany}
}
\end{center}

\begin{abstract}
\noindent
We point out the importance of the decay channels $A\to Zh$
and $H\to VV$ in the wrong-sign limit of the Two-Higgs-Doublet Model (\thdm{}) of type II.
They can be the dominant decay modes at moderate values of $\tan\beta$,
even if the (pseudo)scalar mass is above the threshold
where the decay into a pair of top quarks is kinematically open.
Accordingly, large cross sections $pp\to A\to Zh$ and $pp\to H\to VV$ are obtained
and currently probed by the \lhc{} experiments, yielding conclusive statements
about the remaining parameter space of the wrong-sign limit.
In addition, mild excesses -- as recently found in the \atlas{} analysis $b\bar b \to A\to Zh$ --
could be explained. The wrong-sign limit makes other important testable
predictions for the light Higgs boson couplings.
\end{abstract}

\section{Introduction}

After the discovery of a Standard Model (\sm{})-like Higgs boson with a mass of $125$\,GeV at the
\lhc{}~\cite{Aad:2012tfa,Chatrchyan:2012xdj} the experimental collaborations put a large effort on the precise
determination of its mass~\cite{Aad:2015zhl} and its couplings to the other \sm{} fermions and gauge
bosons~\cite{Khachatryan:2016vau}.
Even though no significant deviations from the \sm{} predictions were observed, there is plenty of room
for the \sm{}-like Higgs boson to be part of an extended Higgs sector.
Accordingly, apart from the precise determination of the \sm{}-like Higgs boson's properties, the
search for additional Higgs bosons is ongoing.
A well-motivated, simple extension of the \sm{} Higgs sector is the class of Two-Higgs-Doublet Models
(\thdm{}s), which were introduced by Lee in 1973~\cite{Lee:1973iz} to explain the observed
matter-antimatter asymmetry as arising from a spontaneous breaking of \cp{}.
The field content of the \sm{} is complemented with an extra $SU(2)$ doublet, which even in
the \cp{}-conserving case results in a richer scalar spectrum: two \cp{}-even scalars,
the lightest $h$ and the heavier Higgs boson~$H$, a pseudoscalar~$A$ and two charged scalars~$H^\pm$.
In the \thdm{} a softly broken $\mathbb{Z}_2$ symmetry is commonly extended to the Yukawa sector
to prevent the appearance of tree-level flavor changing neutral currents~\cite{Glashow:1976nt, Paschos:1976ay}.
This procedure results in four distinct \thdm{} types, some of which include a peculiar and interesting
region of parameter space called the wrong-sign
limit~\cite{Ferreira:2014naa,Ferreira:2014dya,Dumont:2014wha,Fontes:2014tga,Bernon:2014nxa,Biswas:2015zgk,
Modak:2016cdm}.
In this region, there is a relative sign between the couplings of the \sm{}-like Higgs boson to down-type
quarks and to gauge bosons with far-reaching phenomenological consequences. This sign change -- the down-type
quark couplings to the \sm{}-like Higgs boson in the 2HDM acquire the opposite sign as in the \sm{} -- is a
physical quantity, as it has physical consequences which cannot be removed by some field redefinition.

Standard searches for additional Higgs bosons focus on the decays into fermions and gauge bosons.
In particular in some \thdm{} types, where the coupling to down-type fermions is enhanced through large
values of $\tan\beta$, searches in the $\tau^+\tau^-$ final state~\cite{Aaboud:2017sjh,CMS-PAS-HIG-16-037}
are very powerful in setting limits on the \thdm{} parameter space.
On the other hand, for low values of $\tan\beta$, the \thdm{} parameter space is probed by heavy Higgs
decays into a pair of top quarks~\cite{Aaboud:2017hnm} or gauge
bosons~\cite{Aaboud:2017fgj,Aaboud:2017itg,Aaboud:2017gsl,ATLAS-CONF-2016-079,ATLAS-CONF-2017-058,
CMS-PAS-HIG-16-001,CMS-PAS-HIG-16-033,CMS-PAS-HIG-16-034}.
Searches in the di-photon channel are especially useful to constrain pseudoscalars below
$m_A<350$\,GeV~\cite{Aaboud:2017yyg, Khachatryan:2016yec}.
Since the discovery of a light Higgs~$h$ at $125$\,GeV  the search for pseudoscalars decaying into a
gauge boson $Z$ and the light Higgs~$h$ has been deserving special attention and was carried out at
$8$\,TeV~\cite{Khachatryan:2015lba,Aad:2015wra} and $13$\,TeV~\cite{ATLAS-CONF-2016-015,ATLAS:2017nxi}.
If the mass splitting between the pseudoscalar and the heavy \cp{}-even Higgs boson is large enough, also
$H\to ZA$ and $A\to HZ$ constrain parts of the parameter space~\cite{Khachatryan:2016are, CMS-PAS-HIG-16-010}.
Lastly, \cp{}-even Higgs decays into a pair of light Higgs bosons probe corners of the parameter
space~\cite{ATLAS-CONF-2016-004,CMS-PAS-HIG-17-008,Aaboud:2016xco,CMS-PAS-B2G-16-026,ATLAS-CONF-2016-049,
ATLAS-CONF-2016-071,Sirunyan:2017guj,Sirunyan:2017djm}.
\citere{Han:2017pfo} provides a recent overview of excluded parameter regions from the various mentioned
decays in the \thdm{}.
In this paper, we emphasize that in the wrong-sign limit of the \thdm{} of types II/F both $A\to Zh$ and
$H\to VV$ can be dominant, even if the decay into a pair of top quarks is kinematically open.
Therefore the small deviation in the search $b\bar b\to A\to Zh$ with $h\to b\bar b$ at masses of the
pseudoscalar around $\sim 440$\,GeV observed by the \atlas{} collaboration, see \citere{ATLAS:2017nxi}, with
$0.1-0.3$\,pb above the \sm{} background, can be explained in the wrong-sign limit.
The search assumed leptonic decays of the $Z$ boson as hadronic decays of the $Z$ boson only become
useful at higher invariant masses of the pseudoscalar~\cite{Aaboud:2017ahz}.
We note that the region of interest was also probed by the $8$\,TeV \atlas{} and \cms{}
analysis~\cite{Khachatryan:2015lba,Aad:2015wra}, which already set bounds on cross sections of similar size.
The initial $13$\,TeV analysis~\cite{ATLAS-CONF-2016-015} carried out by the \atlas{} collaboration with an
integrated luminosity of $3.2$fb$^{-1}$ also shows upward fluctuations, though at larger cross sections.
If the small excess is confirmed, a signal in the gluon-fusion induced production mode is likely to be seen as
well, at least at intermediate values of $5<\tan\beta<7.5$.
On the other hand, if not confirmed, the upcoming data accumulated in the searches for heavy Higgs bosons,
both in $A\to Zh$ and $H\to VV$, will conclusively probe the remaining parameter space of the wrong-sign limit.
In contrast, the decay of the \cp{}-even Higgs boson into two light \sm{}-like Higgs bosons is not yet
sensitive to the parameter region under discussion.

The wrong-sign limit also makes testable predictions for the light Higgs boson couplings, most prominently
it enhances the gluon-fusion cross section and simultaneously reduces the partial width into two photons.
These deviations from the \sm{} Higgs properties of the light Higgs boson do not decouple and therefore also
limit the parameter space of the wrong-sign limit.
We base our numerical analysis on a data set that does not only take into account the light Higgs boson data
and the searches for heavy Higgs bosons, but also theoretical considerations like boundedness from below,
stability and perturbativity of the scalar potential.
It also respects bounds from $B$-physics and electroweak precision measurements.

Our paper is organized as follows: In \sct{sec:2hdm} we provide an introduction to the Two-Higgs-Doublet
Model including the four different types and define the wrong-sign limit.
Based on an extensive data set generated for the \thdm{} of type II, \sct{sec:ppAZh} explains the enhancement
of $pp\to A\to Zh$ in the wrong-sign limit.
We discuss other phenomenological consequences of the wrong-sign limit in \sct{sec:otheraspects}. In particular,
we address the enhancement of $pp\to H\to VV$, which is also under current experimental investigation and
present the deviations in the light Higgs boson signal strengths induced by the wrong-sign limit.
We conclude in \sct{sec:conclusions}.\\

\section{The \thdm{} and the wrong-sign limit}
\label{sec:2hdm}

As we argued in the introduction, the Two-Higgs-Doublet Model (\thdm{}) is one of the simplest extensions of the \sm{} Higgs sector.
Despite its simplicity, the model boasts a rich phenomenology.  Spontaneous \cp{} breaking is possible, some versions of the model provide natural candidates for dark matter or tree-level flavor changing neutral currents (FCNC) whose magnitude can be made naturally small via appropriate symmetries (for a recent review, see \cite{Branco:2011iw}).
In the most used versions of the model,  a global $\mathbb{Z}_2$ symmetry is imposed on the Lagrangian, softly broken by a dimension two term.
This symmetry is introduced to eliminate tree-level flavor changing neutral currents in the Yukawa sector~\cite{Glashow:1976nt, Paschos:1976ay}.
The scalar potential for this version of the model may be written as
\begin{align}
 V &= m^2_{11}|\Phi_1|^2+m^2_{22}|\Phi_2|^2
 +m^2_{12}\left[\Phi_1^\dagger\Phi_2+{\rm h.c.}\right] \nonumber \\
  &\hphantom{=}+\frac{\lambda_1}{2}|\Phi_1|^4
   +\frac{\lambda_2}{2}|\Phi_2|^4 +\lambda_3 |\Phi_1|^2 |\Phi_2|^2   +\lambda_4 |\Phi_1^\dagger\Phi_2|^2
   +\frac{\lambda_5}{2}\left[\left(\Phi_1^\dagger\Phi_2\right)^2+{\rm h.c.}\right]\,,
\label{eq:pot}
\end{align}
with all $8$ parameters real.
The quartic couplings of the model must obey well-known bounded from below conditions~\cite{Deshpande:1977rw,Klimenko:1984qx,Ivanov:2006yq,Ivanov:2007de} and other
constraints to ensure perturbative unitarity \cite{Kanemura:1993hm,Akeroyd:2000wc}.

After spontaneous symmetry breaking the doublets $\Phi_1$ and $\Phi_2$ acquire real vacuum expectation
values (vevs), $v_1$ and $v_2$.
To ensure that the electroweak gauge bosons have their known masses the relation $v_1^2 + v_2^2 = v^2$ with
$v = 246$\,GeV, has to hold.
The model's scalar sector is usually described in terms of the following set of parameters: the vev $v$;
the four physical scalar masses (those of $h$, $H$, $A$ and $H^\pm$); the angle $\beta$, defined such that
$\tan\beta = v_2/v_1$; the angle $\alpha$, which is the diagonalization angle for the $(2\times 2)$
\cp{}-even mass matrix; and the soft-breaking term in the scalar potential, $m^2_{12}$.
Without loss of generality, we may take the interval of variation of $\alpha$ to be
 $-\pi/2 \leq \alpha \leq \pi/2$.
Notice both possible signs for this parameter.
In all that follows, we will assume that the light \cp{}-even Higgs boson~$h$ is the one that has
been observed at the \lhc{}. Thus we fix $m_h = 125$ GeV.

As for the Yukawa sector, it can be shown that for the $\mathbb{Z}_2$ symmetry to eliminate
tree-level FCNC, each set of same-charge fermions should couple to a single scalar doublet.
This leaves four possibilities for the scalar-fermion couplings, summarized in \tab{tab:types}.
\begin{table}
\begin{center}
\begin{tabular}{rccc} \toprule
& $u$-type & $d$-type & leptons \\ \midrule
Type I & $\Phi_2$ & $\Phi_2$ & $\Phi_2$ \\
Type II & $\Phi_2$ & $\Phi_1$ & $\Phi_1$ \\
Lepton Specific (LS) & $\Phi_2$ & $\Phi_2$ & $\Phi_1$ \\
Flipped (F) & $\Phi_2$ & $\Phi_1$ & $\Phi_2$ \\ \bottomrule
\end{tabular}
\caption{Couplings between fermion and Higgs doublets in the four Yukawa types of the softly
broken $\mathbb{Z}_2$-symmetric \thdm{}. \label{tab:types}}
\end{center}
\end{table}
In what regards the quark sector, models I and Lepton Specific (LS) have almost identical phenomenologies, as do models II and Flipped (F).
There are significant constraints on the models' parameter space, stemming from $B$-physics, in
particular from $b\rightarrow s\gamma$ measurements~\cite{Deschamps:2009rh,Mahmoudi:2009zx,Hermann:2012fc,
Misiak:2015xwa,Misiak:2017bgg}.
Roughly, these constraints translate into a lower bound on $\tan\beta$ (we will take $1 \leq \tan\beta\leq 35$)
and, for Model II and the Flipped model, a lower bound on the mass of the charged scalar (we will take $m_{H^\pm}
\geq 480$ GeV).
Please notice that the latest results from~\citere{Misiak:2017bgg} already put this bound at about
$580$\,GeV, a point to which we shall return.

The mixing between the two Higgs doublet alters the couplings of $h$ {\em vis a vis} what one would expect
if the lightest \cp{}-even was exactly the \sm{} Higgs.
It is customary to define
\begin{align}
\kappa_X^2=\frac{\Gamma^{\scriptscriptstyle {\rm \thdm{}}}  (h \to X)}{\Gamma^{\scriptscriptstyle {\rm \sm{}}} (h \to X)}
\end{align}
for each \sm{} decay state $X$.
At tree-level, this ratio is simply the square of the ratio of the couplings
$\kappa_X= g_{X}^{\scriptscriptstyle {\rm \thdm{}}}   /g_{X}^{\scriptscriptstyle {\rm \sm{}}} $.
For the $ZZ$ and $WW$ couplings to $h$, one obtains
\begin{align}
\kappa_{VV} \,=\, \sin(\beta - \alpha)
\end{align}
for all \thdm{} model types.
Given the ranges allowed for both $\alpha$ and $\beta$ and \lhc{} results indicating \sm{}-like behavior for $h$,
one finds that $\kappa_{VV}\,>\,0$, so the
coupling of $h$ to the electroweak gauge bosons in the \thdm{} always has the same sign as in the \sm{}.
As for the couplings of $h$ to up-type quarks, they are such that
\begin{align}
\kappa_U\,=\, \frac{\cos\alpha}{\sin\beta}
\end{align}
for all Yukawa models, since $\Phi_2$ always couples to up-type quarks.
For down-type quarks, one has for models I and LS
\begin{align}
\kappa_D\,=\, \frac{\cos\alpha}{\sin\beta}
\label{eq:kd}
\end{align}
and for models II and F,
\vspace{-5mm}
\begin{align}
\kappa_D\,=\, -\,\frac{\sin\alpha}{\cos\beta}\,.
\end{align}
Given that $\tan\beta>0$ and the angle $\alpha$ ranges from $-\pi/2$ to $\pi/2$,  it is easy to see that $\kappa_U>0$.
However, though for models I and LS $\kappa_D>0$, for models II and F that is no longer true --- the coupling of the light Higgs boson~$h$ to down-type quarks in the \thdm{} of type II/F can have the opposite sign to the respective \sm{} coupling and to $\kappa_{VV}$.
This {\em wrong-sign limit}
\cite{Ferreira:2014naa,Ferreira:2014dya,Dumont:2014wha,Fontes:2014tga,Bernon:2014nxa,Biswas:2015zgk,Modak:2016cdm}
is realized if  $\alpha \,>\,0$. More useful information can be gathered if one rewrites~\eqn{eq:kd} as
\begin{align}
\kappa_D\,=\, -\,\frac{\sin\alpha}{\cos\beta}\,=\,\sin(\beta - \alpha) \,-\,
\cos(\beta - \alpha)\,\tan\beta\,.
\end{align}
Since the light Higgs boson~$h$ should behave like the \sm{} Higgs boson~\cite{Khachatryan:2016vau} one must have $\sin(\beta - \alpha) \simeq 1$.
Thus the only possibility to have a negative sign in $\kappa_D$ is if $\cos(\beta - \alpha)\,\tan\beta \simeq 2$, which points to higher values of $\tan\beta$  and $\cos(\beta - \alpha)\,>\,0$.
In fact, current \lhc{} results still allow for sizeable values of $\cos(\beta - \alpha)$ ($\sim  0.3-0.4$).

The wrong-sign limit is not merely a strange corner of parameter space. It has, in fact, phenomenological
consequences observable at the \lhc{}.
The change in sign of the bottom-quark coupling affects both the gluon-fusion production cross section --
enhancing it -- and the di-photon branching ratio of the light Higgs boson~$h$ -- suppressing it.
\footnote{The enhancement of the gluon fusion cross section can be understood by expanding it in terms of the dominant top- and bottom-quark amplitudes, $A_t$ and $A_b$.
The cross section is proportional to $|A_t+A_b|^2=|A_t|^2+\text{Re}(A_tA_b)+|A_b|^2$.
The second term changes sign upon flipping the sign of the bottom-quark Yukawa coupling.
Since this term is negative in the \sm{} this leads to an enhancement of the cross section in the wrong-sign limit (see also footnote~\ref{foot:f}).
A similar argument holds for the di-photon branching ratio with
additional $W$ boson and charged Higgs boson in the loop, where
the effect of the negative bottom quark Yukawa is however
much less pronounced.
The dominant negative interference for the di-photon
branching ratio is induced through the
additional charged Higgs boson, see the discussion of \fig{fig:ggZZ}.}
As a consequence, the wrong-sign limit is an example of a nondecoupling regime within the \thdm{}, as
it cannot lead to light Higgs boson properties exactly equal to ones of the \sm{} Higgs
boson~\cite{Ferreira:2014naa}, a point which we will address again in \sct{sec:otheraspects}.
We note that the wrong-sign limit is also under tension from requesting validity of the model up to
just a few TeV~\cite{Basler:2017nzu}.
We will now consider the effect of the wrong-sign limit on other observables, in particular the production
of a pseudoscalar and its subsequent decay to $Zh$.
In the \thdm{} of type II the couplings $A-Z-h$ and $H-V-V$ are proportional to $\cos(\beta-\alpha)$ and
therefore lead to sensitivity to the wrong-sign limit in searches for heavy scalars in $Zh$ and $VV$ final states.
Lastly, the coupling $H-h-h$ also increases with $\cos(\beta-\alpha)$ yielding a potential sensitivity
in future di-Higgs searches.

\section{$pp \to A\to Zh$ in the wrong-sign limit}
\label{sec:ppAZh}

Recently, in \citere{ATLAS:2017nxi}, the \atlas{} collaboration reported a small deviation on the search channel
$pp \to A\to Zh$: an excess, relative to background expectations, of $0.1-0.3$\,pb for $\sigma(pp \to A\to Zh)
\text{BR}(h\to b\bar{b})$, for a potential pseudoscalar mass of about $440$\,GeV.
The significance is larger in case the pseudoscalar~$A$ is produced through bottom-quark annihilation rather
than gluon fusion, but both production processes show a deviation.
The statistical significance of this excess is (yet) too low to indicate anything meaningful, but it does raise
the question: would it be possible to account for such an excess within the framework of the \thdm{}?
Common wisdom suggests that for that mass range both the production cross sections and the branching ratios
into $Zh$ would be too small for an excess to occur.
In fact, for such masses, one is above the $t\bar{t}$ threshold and the cross section for gluon fusion is expected to decrease rapidly.
Furthermore, since the pseudoscalar can decay to a pair of top quarks, that decay channel would be expected to dominate over all others.

\begin{figure}
\begin{center}
\includegraphics[height=8cm,angle=0]{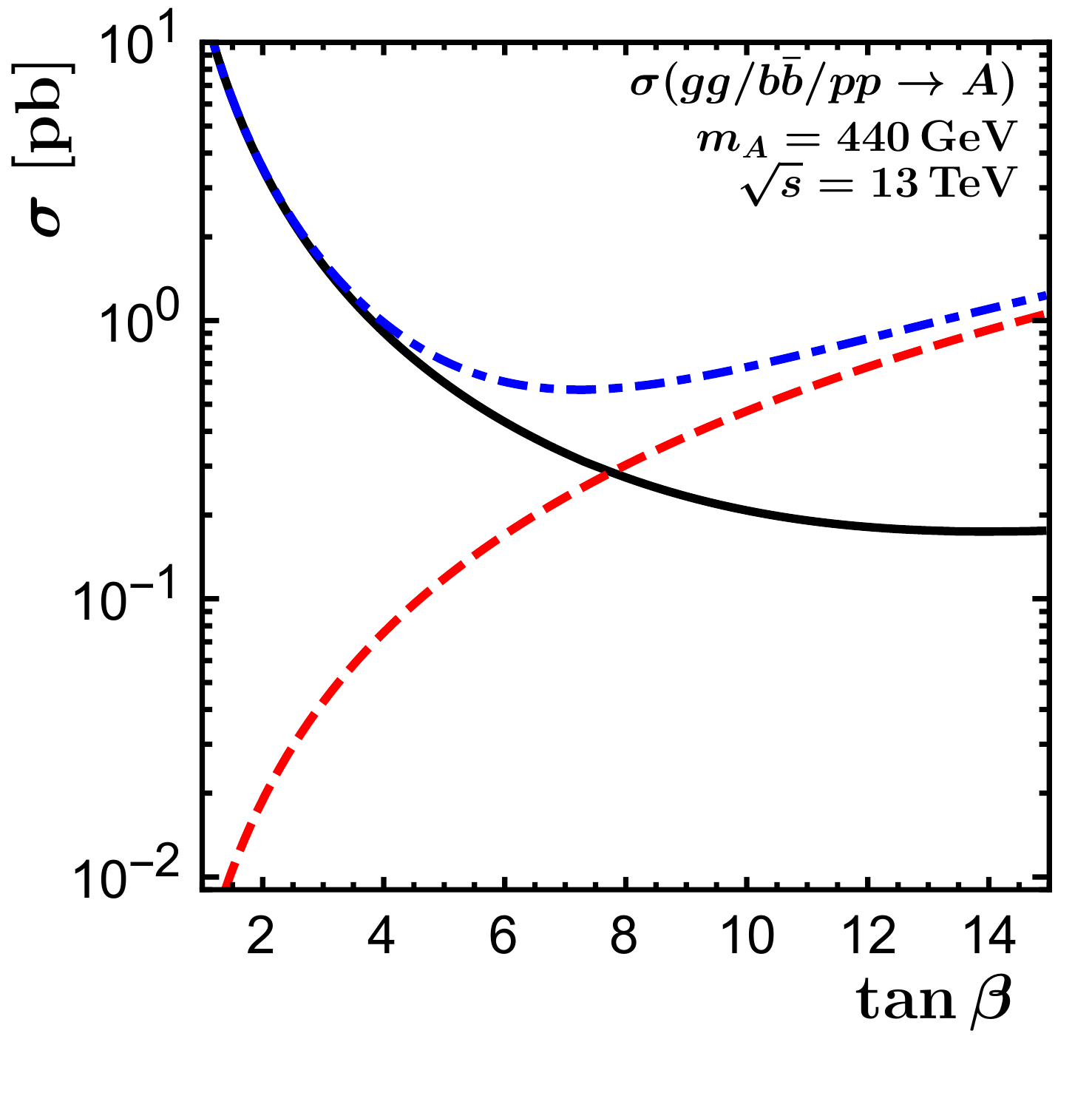}
\end{center}
\vspace{-5mm}
\caption{Gluon-fusion cross section (black, solid) and bottom-quark annihilation cross section (red, dashed)
and their sum (blue, dot-dashed) for a pseudoscalar with mass $440$\,GeV at a center-of-mass energy of $13$\,TeV
as a function of $\tan\beta$.
}
\label{fig:ggAbbA}
\end{figure}

However, several details of the \thdm{} allow to overcome those initial difficulties. To wit:
\begin{itemize}
\item The gluon-fusion production cross section is expected to be larger for a pseudoscalar~$A$ (of a \thdm{} with $\tan\beta=1$) than for a \sm{} scalar~$H$.
At leading order, both processes occur due to a triangle fermion loop diagram.
To illustrate the enhancement for a pseudoscalar, we compare the top-triangle contribution to the production cross section for a pseudoscalar mass of $m_A = 440$ GeV with the analogous contribution for a scalar of the same mass.
We conclude that the pseudoscalar cross section is $|a^A_q(x_A)/a^H_q(x_H)|^2\,\simeq\,2.36$ times larger than the scalar cross section, with $x_\phi = m_\phi^2/4 m_t^2$.
The  functions~$a_q^\phi$ can e.g. be found in \citere{Harlander:2012pb}. The ratio is also hardly affected by higher-order contributions.
At next-to-next-to-leading order (\nnlo{}) \qcd{}~\cite{Harlander:2002wh,Anastasiou:2002yz,Ravindran:2003um,Harlander:2002vv,Anastasiou:2002wq} it yields $\sigma(gg\to A)/\sigma(gg\to H)=17.03\,\text{pb}/7.16\,\text{pb}\sim 2.38$ at a center-of-mass energy of $13$\,TeV even including bottom- and charm-quark effects at next-to-leading (\nlo{}) \qcd{}~\cite{Spira:1995rr}.
The numbers are produced with \sushi{1.6.0}~\cite{Harlander:2012pb,Harlander:2016hcx}.
\item In the \thdm{} of type II, the coupling of the pseudoscalar to the top quark is proportional to $1/\tan\beta$, but the coupling to bottom quarks grows with $\tan\beta$.
Therefore, with increasing $\tan\beta$ bottom-quark annihilation $b\bar b\to A$ is a sizeable contribution to pseudoscalar Higgs production $pp\to A$.
Traditionally, it can be described at \nnlo{} \qcd{} in the five-flavor scheme~\cite{Harlander:2003ai} or at \nlo{} \qcd{} in the four-flavor scheme~\cite{Dittmaier:2003ej,Dawson:2003kb} and theory predictions were based on cross sections matching qualitatively between the two schemes~\cite{Harlander:2011aa}.
Since recently theoretically clean matching procedures are also available~\cite{Bonvini:2015pxa,Bonvini:2016fgf,Forte:2015hba,Forte:2016sja}, which are numerically very close to the five-flavor
scheme.
We thus proceed with bottom-quark annihilation described in the five-flavor scheme as implemented in \sushi{1.6.0}.
We compare the gluon-fusion and the bottom-quark annihilation cross sections in the five-flavor scheme for a pseudoscalar of $m_A=440$\,GeV at $13$\,TeV as a function of $\tan\beta$ in \fig{fig:ggAbbA}.
Around $\tan\beta=7.5$ both gluon fusion and bottom-quark annihilation are of a similar size, namely $0.3$\,pb each.
For slightly larger values of $\tan\beta$ $b\bar b\to A$ is a bit larger than $gg\to A$, which appears to be the case for the mild excess in \citere{ATLAS:2017nxi}.
\item The coupling of the pseudoscalar to $Zh$ is proportional to $\cos(\beta-\alpha)$
which, in the wrong-sign limit, can have values of the order of $0.2-0.6$. Though the
branching ratio BR$(A\to Zh)$ will be proportional to the square of this number, the
magnitude of it is sufficient to make this the preferred decay channel of $A$ for
certain regions of the parameter space, namely those which include the wrong-sign limit.
In that case, for intermediate values of $\tan\beta$, the decay mode into $Zh$ also dominates
over the decay modes into quark pairs, both $t\bar t$ and $b\bar b$.
\end{itemize}

We emphasize that within the Minimal Supersymmetric Standard Model (\mssm{}) such large values of $\cos(\beta-\alpha)$ are
hard to obtain, in particular since in the decoupling limit with $m_A\gg m_Z$
$\cos(\beta-\alpha)\to 0$ is quickly reached. The alignment limit
usually occurs at larger values of $\tan\beta$~\cite{Carena:2013ooa},
if not large values for the soft-breaking parameters or $\mu$ are chosen.

To illustrate the above statements we use a data set generated by {\tt ScannerS}~\cite{Coimbra:2013qq,Ferreira:2014dya,Costa:2015llh,Muhlleitner:2016mzt}.
We imposed the tree-level theoretical constraints of boundedness from below~\cite{Klimenko:1984qx,Branco:2011iw} and perturbative unitarity~\cite{Branco:2011iw}.
We additionally required the electroweak vacuum to be the global minimum of the tree-level Higgs potential~\cite{Barroso:2013awa}.
These constraints were imposed only at the electroweak scale. We did not consider the issue of requiring validity of the
model up to higher scales, which would further constrain the theory's parameter space
(see, for instance, \citeres{Chakrabarty:2014aya,Chakrabarty:2017qkh,Basler:2017nzu}).
We checked experimental bounds from $B$-physics, of which the aforementioned constraint on $b\rightarrow s\gamma$ is the most important one.
Electroweak precision measurements were taken into account through bounds on the oblique parameters $S$, $T$ and $U$~\cite{Baak:2014ora}.
The branching ratios, decay widths and production cross sections of all Higgs bosons were calculated using
{\tt HDECAY 6.511}~\cite{Djouadi:1997yw,Butterworth:2010ym} and \sushi{1.6.0}~\cite{Harlander:2012pb,Harlander:2016hcx}\footnote{For the parameter regions of interest these
results were verified with {\tt HIGLU}~\cite{Spira:1995mt} and {\tt \thdm{}C}~\cite{Eriksson:2009ws}, respectively. We refer to
\citere{Harlander:2013qxa} for a comparison of the numerical codes available for the \thdm{}.}. Thus, the bottom-quark annihilation cross section
is based on the five-flavor scheme description.
This information was  passed to
{\tt HiggsBounds 4.3.1}~\cite{Bechtle:2008jh,Bechtle:2011sb,Bechtle:2013wla,Bechtle:2015pma} to check exclusion bounds from searches for additional Higgs bosons.
We finally required that the 125 GeV scalar is very much \sm{}-like. We achieve this
by demanding that the rates
\begin{align}
\mu_X\,=\,\frac{\sigma^{\rm \thdm{}}(pp\to h)\,\text{BR}^{\rm \thdm{}}(h\to X)}{\sigma^{\rm \sm{}}(pp\to h)\,\text{BR}^{\rm \sm{}}(h\to X)}
\end{align}
are within at most $20\%$ of their expected \sm{} values ({\em i.e.} $1$). This ensures
a very good compliance with current \lhc{} bounds~\cite{Khachatryan:2016vau}.
As already stated, we allow $-\pi/2\leq \alpha \leq\pi/2$ and $1\leq \tan\beta\leq 35$.
The masses of the non-\sm{}-like Higgs bosons are constrained to $30\,\mathrm{GeV}\leq m_A\leq 1\,\mathrm{TeV}$,
$130\,\mathrm{GeV}\leq m_H\leq 1\,\mathrm{TeV}$ and $480\,\mathrm{GeV}\leq m_{H^\pm}\leq 1\,\mathrm{TeV}$
and the soft $\mathbb{Z}_2$ breaking parameter to $1\,\mathrm{GeV}^2\leq m_{12}^2\leq 5\times10^5\,\mathrm{GeV}^2$.

Before we explain the details of our results, we emphasize that for our purposes
the usage of the narrow-width approximation is still allowed and interference
effects among Higgs bosons and/or the \sm{} background can be neglected:
in the \thdm{} \citere{Greiner:2015ixr} pointed out the importance of
interference effects in $gg(\to H)\to VV$ among the heavy Higgs boson,
the light Higgs boson and the \sm{} background. Three reference
points include a heavy Higgs mass of $400$\,GeV, very close to
the mass region discussed here. For those points large interference occurs in
parameter regions where the signal contribution $gg\to H\to VV$
has a cross section below $\sim 10^{-2}$\,pb for moderate values
of $\tan\beta$, i.e. beyond the interesting parameter
regions discussed here. The same conclusion can be derived
for $b\bar b(\to H)\to VV$.
A similar statement holds for $pp(\to A)\to Zh$, for which both
$gg\to A\to Zh$ and $b\bar b\to A\to Zh$ including the interferences
with the background were discussed in \citere{Harlander:2013mla}.
A new version of {\tt vh@nnlo}~\cite{Harlander:2018yio} will soon allow to study
such interferences, at least at leading order in perturbation theory.
Therefore, for all relevant processes in this paper, $gg/b\bar b\to H\to VV$ and
$gg/b\bar b\to A\to Zh$, we can rely on the narrow-width approximation. This allows to take into account higher-order contributions both in production and decay separately.

\begin{figure}[t]
\begin{tabular}{cc}
\includegraphics[height=6cm,angle=0]{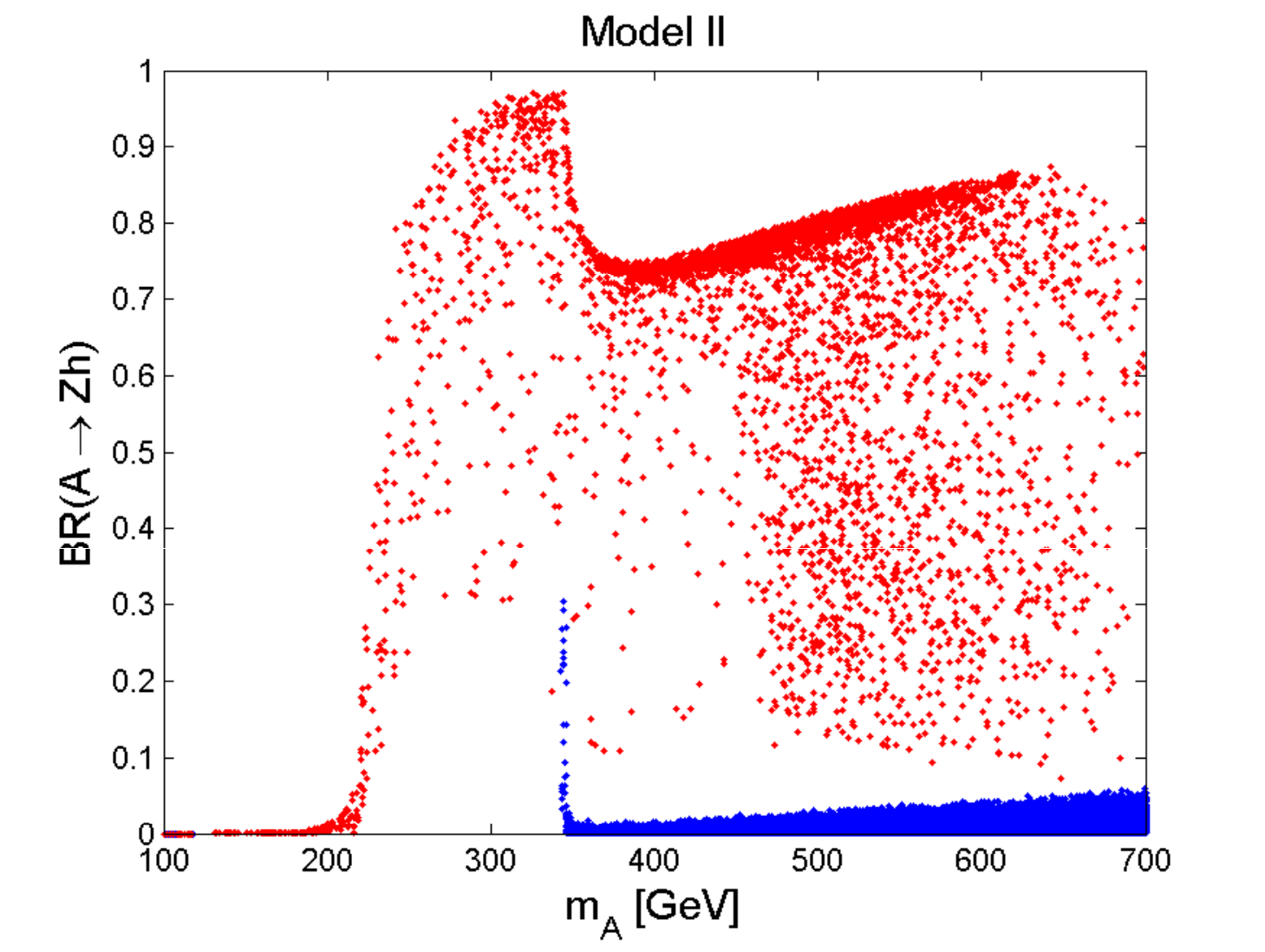}&
\includegraphics[height=6cm,angle=0]{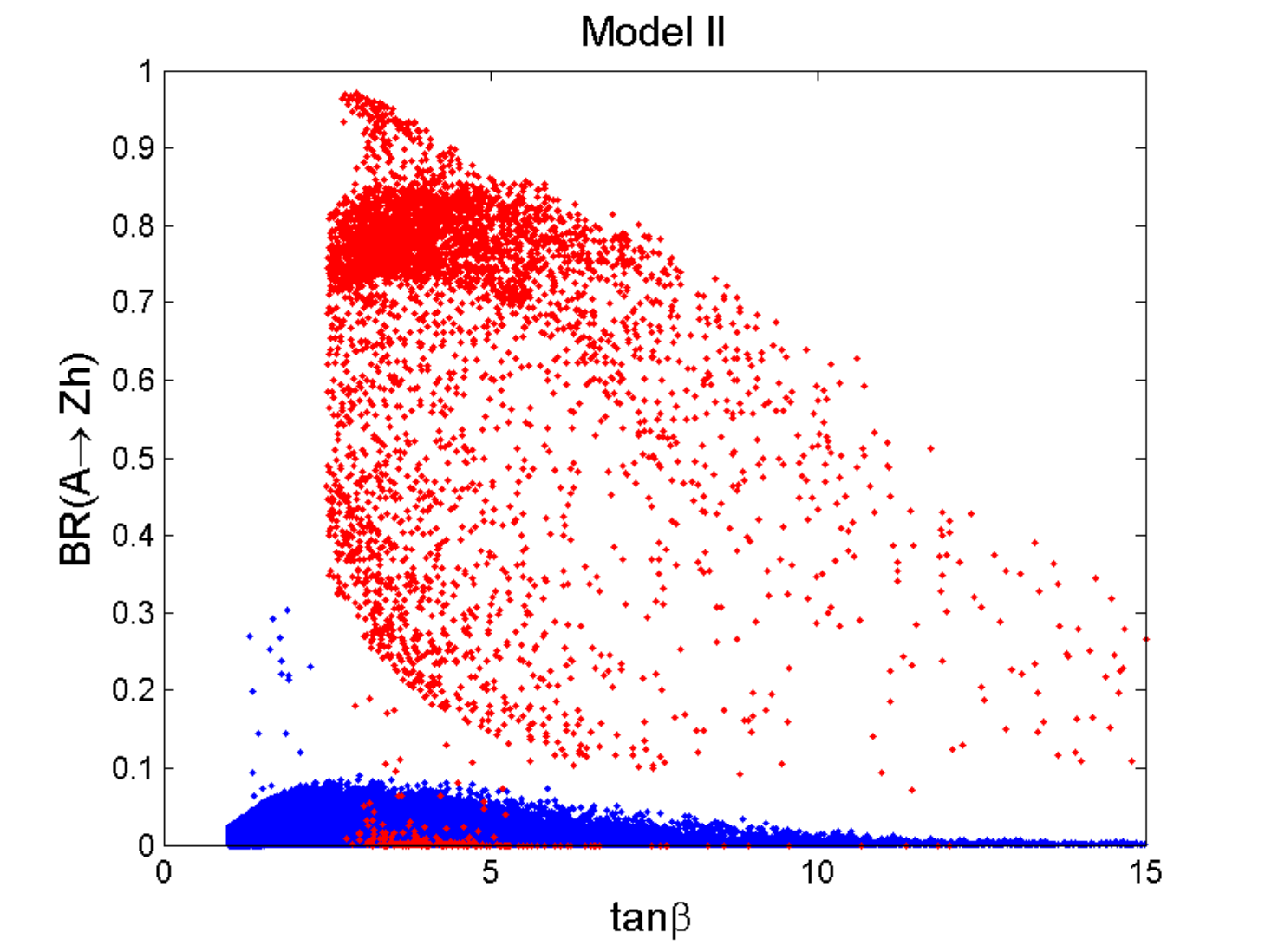}\\
 (a) & (b)
\end{tabular}
\caption{Scatter plots for the \thdm{} of type II showing
(a) BR$(A \to Zh)$ as a function of the pseudoscalar mass $m_A$ in GeV and
(b) BR$(A\to Zh)$ as a function of $\tan\beta$.
All relevant constraints are imposed and $h$ rates are within $20\%$ of their \sm{} values.
Red points are in the wrong-sign limit.
}
\label{fig:crbr}
\end{figure}

We first show the branching ratio of $A\to Zh$ in
\fig{fig:crbr} as a function of the pseudoscalar mass $m_A$ and $\tan\beta$.
We highlighted, in red, the regions of parameter
space corresponding to the wrong-sign limit. It is clear from the figures that
in the wrong-sign limit the decay $A\to Zh$ can indeed become the dominant one, even when the \thdm{} parameter
space is constrained by the \lhc{} requirement that the lightest \cp{}-even Higgs boson~$h$ is \sm{}-like.
\fig{fig:crbr}~(a) demonstrates that large branching ratios are also obtained
above the threshold where the decay into a top-quark pair is kinematically open.
\fig{fig:crbr}~(b) shows that the largest branching ratios are obtained
for $\tan\beta$ within $3$ to $7$. On the one hand, for moderate values
of $\tan\beta$ the decay into a top-quark pair is suppressed and the decay
into a bottom-quark pair is not yet sufficiently enhanced.
The minimum of the sum of the partial decay widths
to $t\bar t$ and $b\bar b$ is at $\tan\beta\sim 7.5$ for $m_A=440$\,GeV.
On the other hand, the wrong-sign limit can accommodate smaller values of $\cos(\beta-\alpha)$ with
increasing $\tan\beta$, which explains why BR$(A\to Zh)$ is reduced with increasing
$\tan\beta$.

\begin{figure}[t]
\begin{tabular}{cc}
\includegraphics[height=6cm,angle=0]{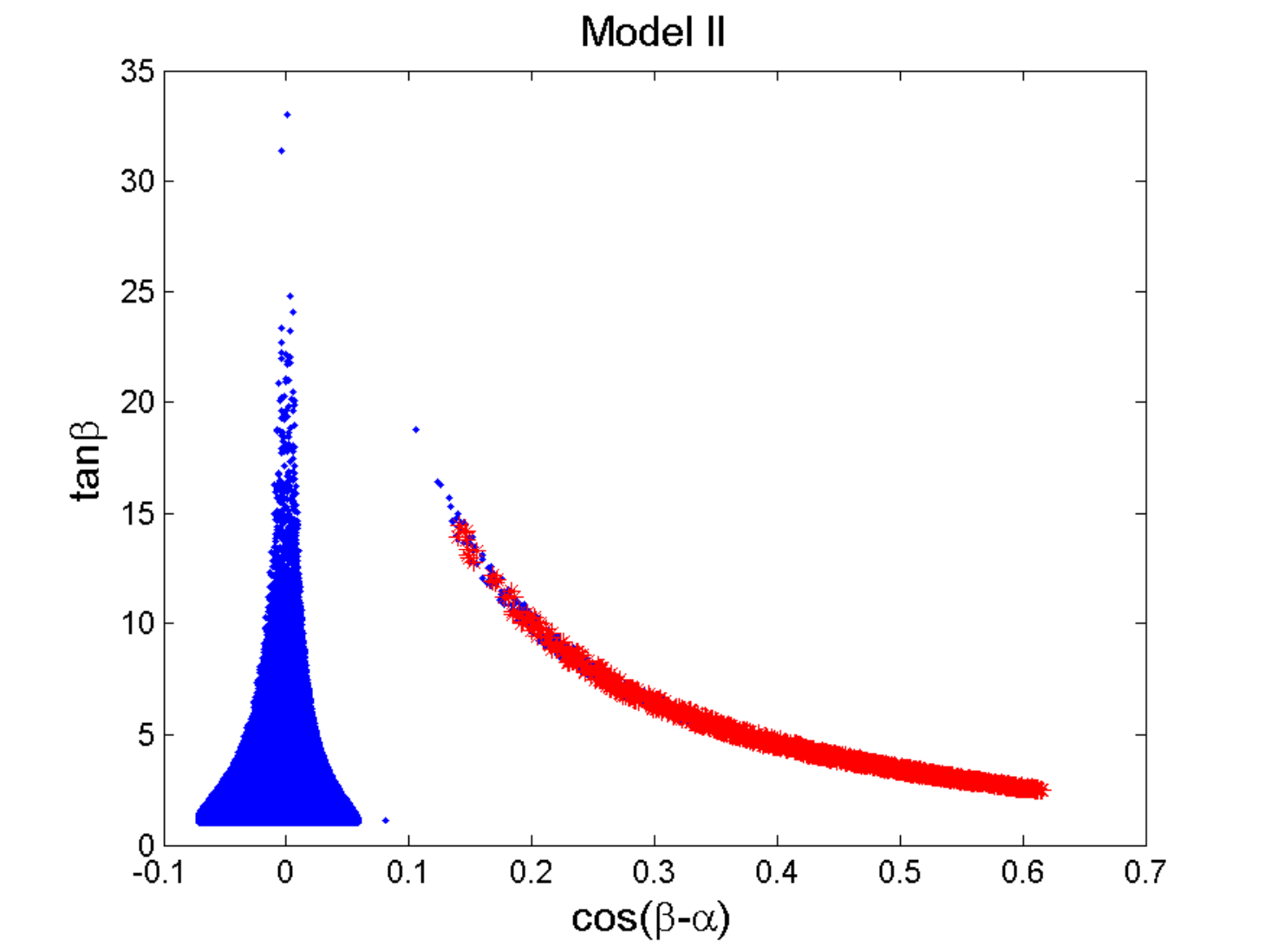}&
\includegraphics[height=6cm,angle=0]{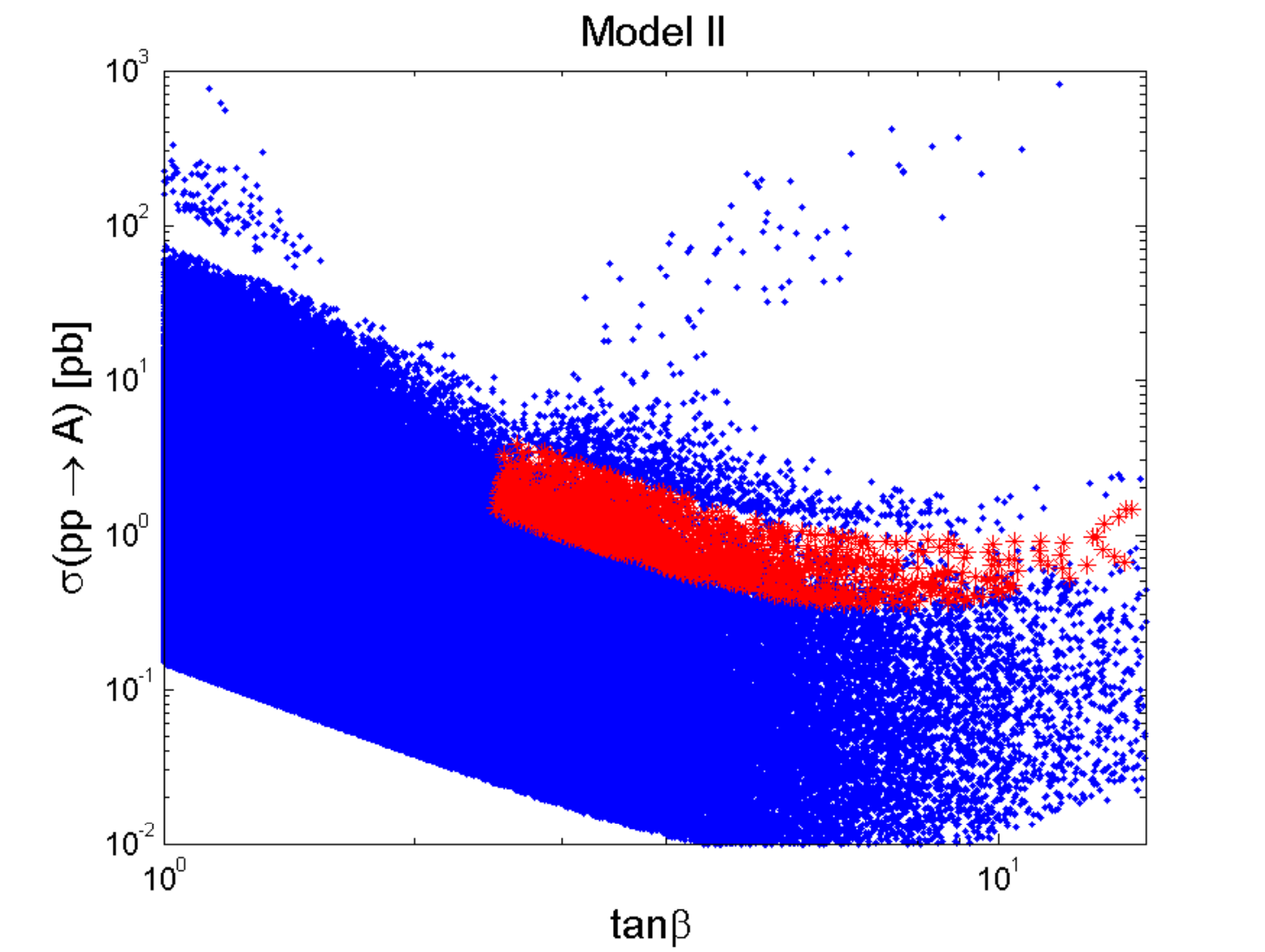}\\
 (a) & (b)
\end{tabular}
\caption{Scatter plots for the \thdm{} of type II showing
(a) $\tan\beta$ versus $\cos(\beta - \alpha)$ and
(b) the cross section for $pp\to A$ in pb as a function of $\tan\beta$.
All relevant constraints are imposed and $h$ rates are within $20\%$ of their \sm{} values.
Red points further satisfy $400\leq m_A\leq 500$ GeV and $\sigma(pp\to A \to Zh\to Zb\bar{b}) \geq 0.1$ pb.
}
\label{fig:ppAZh}
\end{figure}

The reduction of $\cos(\beta-\alpha)$ with increasing $\tan\beta$ is also apparent in \fig{fig:ppAZh}~(a).
The blue points are, as before, those for which the light Higgs boson rates are within $20\%$ of the expected \sm{} values.
In contrast to the previous figures the red points correspond to the subset of the blue points
for which $400\leq m_A\leq 500$ GeV and $\sigma(pp\to A \to Zh \to Zb\bar{b}) \geq 0.1$ pb.
Here we sum up both gluon fusion and bottom-quark annihilation to obtain $\sigma(pp \to A)$.
As explained in~\citere{Ferreira:2014naa}, the curved band on the right of the plot corresponds to the wrong-sign limit
-- and indeed we see all red points lie along this band. Thus we confirm
that an excess observed in $pp\to A\to Zh$ is associated with larger positive values of $\cos(\beta - \alpha)$
and thus the wrong-sign limit.
Finally, \fig{fig:ppAZh}~(b) depicts the production cross section $\sigma(pp\to A)$ in pb
as a function of $\tan\beta$, again summing up both gluon fusion and bottom-quark annihilation.
Notice how the red points, for which the cross section of $pp\to A \to Zh$ is sizeable, seem to follow a descending line up
to $\tan\beta \simeq 7.5$, and increase from that point on. This inflexion marks the value of
$\tan\beta$ for which the bottom-quark initiated production process becomes as important as
gluon fusion, as we illustrated in \fig{fig:ggAbbA}.
\begin{figure}[t]
\begin{tabular}{cc}
\includegraphics[height=6cm,angle=0]{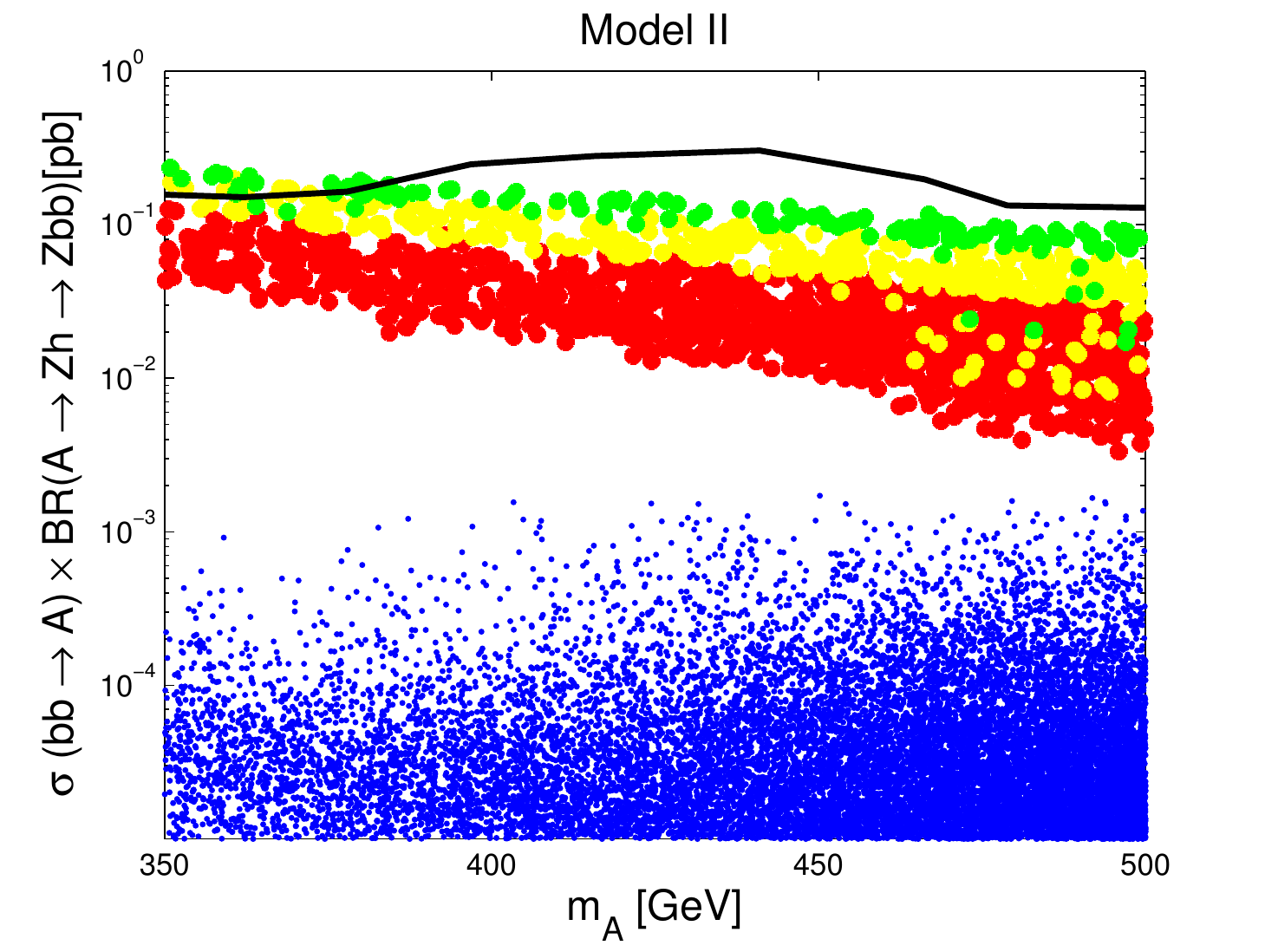}&
\includegraphics[height=6cm,angle=0]{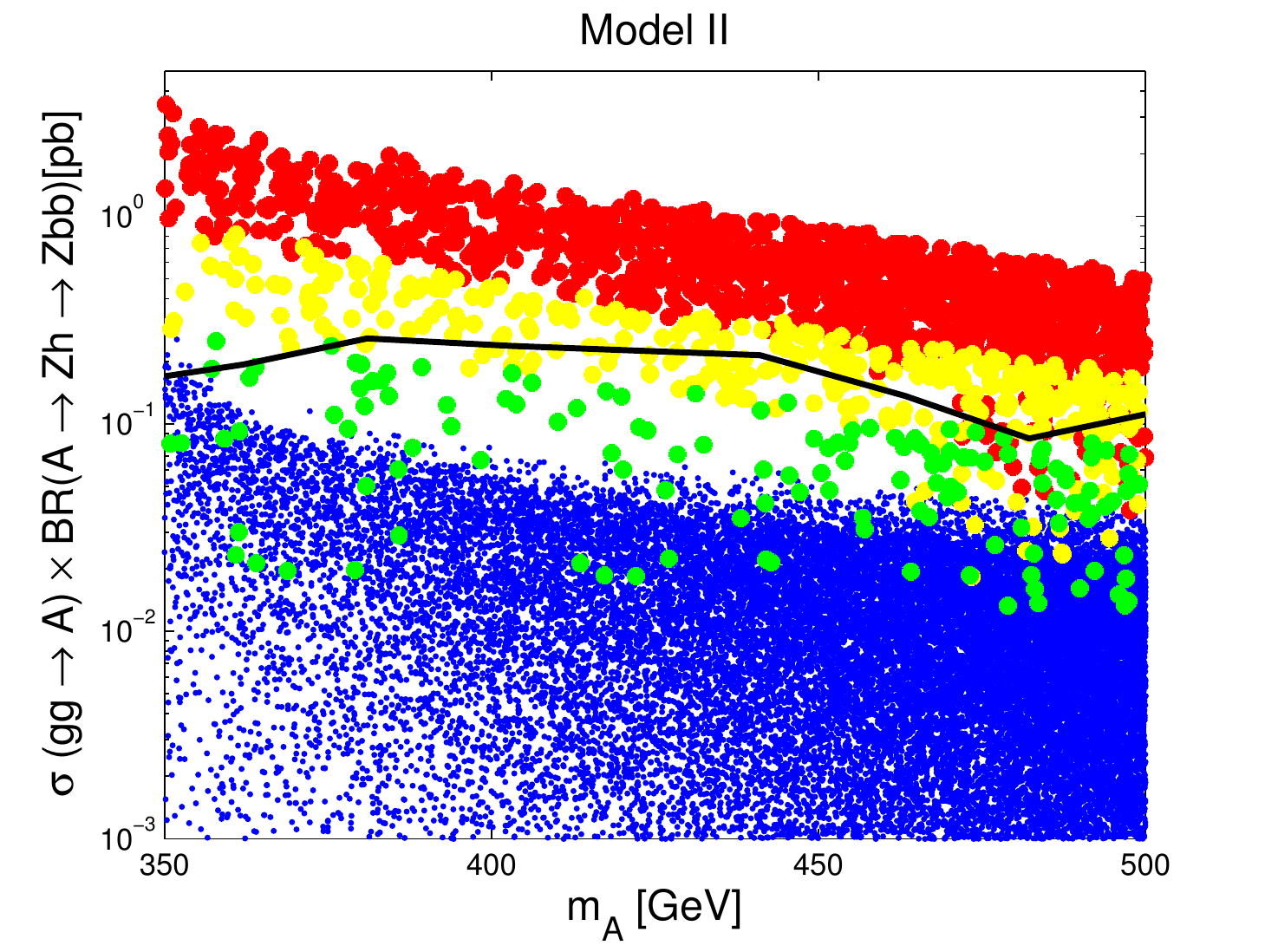}\\
 (a) & (b)
\end{tabular}
\caption{Scatter plots for the \thdm{} of type II showing
(a) the cross section $\sigma(b\bar b\to A){\rm BR}(A\to Zh\to Z b\bar b)$ in pb as a function of $m_A$ in GeV and
(b) the cross section $\sigma(gg\to A){\rm BR}(A\to Zh\to Z b\bar b)$ in pb as a function of $m_A$ in GeV.
All relevant constraints are imposed and $h$ rates are within $20\%$ of their \sm{} values.
Red, yellow and green points represent those parameters satisfying the wrong-sign limit
and distinguish different regions of $\tan\beta$: $1<\tan\beta<5$ (red), $5<\tan\beta<7.5$ (yellow) and $\tan\beta>7.5$ (green).
The black lines indicate the experimental bounds from \citere{ATLAS:2017nxi}.
}
\label{fig:ggbbAZh}
\end{figure}

Finally, we split the cross section into gluon fusion and bottom-quark annihilation and combine
it with the branching ratios $A\to Zh$ and $h\to b\bar b$ in \fig{fig:ggbbAZh}~(a) and (b).
In those figures we also indicate the measured bounds from the \atlas{} analysis~\cite{ATLAS:2017nxi} as black lines.
For the wrong-sign limit we distinguish different regions of $\tan\beta$, namely
$1<\tan\beta<5$, $5<\tan\beta<7.5$ and $\tan\beta>7.5$ with red, yellow and green points, respectively.
It is apparent that in most of the wrong-sign limit parameter space larger cross sections are obtained for
the gluon-fusion process, which can be understood from the fact that points with $\tan\beta<5$
are favoured in our data set, see \fig{fig:crbr}~(b). Still the wrong-sign limit allows for sufficient room to also
have $\sigma(b\bar b\to A){\rm BR}(A\to Zh\to Z b\bar b)>0.1$\,pb at values of $\tan\beta>5$.
Such values of $\tan\beta$ are on the other hand not yet excluded by the gluon-fusion induced process, see \fig{fig:ggbbAZh}~(b).
Please note that a much larger cross section observed in the bottom-quark initiated production process is hard to achieve theoretically.
Thus if the excess in $\sigma(b\bar b\to A){\rm BR}(A\to Zh\to Z b\bar b)$ rises to a significantly larger cross section,
the wrong-sign limit of the \thdm{} of type II cannot provide an explanation.
It is also clear that the investigation
of both production processes separately allows for a conclusive probe of the wrong-sign limit of the \thdm{} of type II.
\citere{Han:2017pfo} shows that at low $\tan\beta$ the process $gg\to A\to Zh$ rules out a large
parameter region of the wrong-sign limit, see e.g. their Fig. 4. This is in full accordance with our observations,
where only values of $\tan\beta>5$ are not yet excluded by $gg\to A\to Zh$.

An interesting feature of the exclusion through the process $pp \to A\to Zh\to Zb\bar b$ is the fact
that $h\to b\bar b$ vanishes for $\cos(\beta-\alpha)\cdot \tan\beta \approx 1$. As indicated in
the caption of Fig. 6 of \citere{ATLAS:2017nxi} this causes a nonexcluded parameter region for low $\tan\beta$
far away from $\cos(\beta-\alpha)=0$. Obviously, this region can however be excluded through the measurement
of the light Higgs boson couplings to bottom quarks and does not correspond to the wrong-sign limit, which occurs
at values of $\cos(\beta-\alpha)\tan\beta$ being twice as large.

\section{Observable consequences of an excess in $pp \to A\to Zh$}
\label{sec:otheraspects}

Since the \thdm{} is such a tightly constrained model, having an excess on a particular
channel usually has consequences for other ones. In the case we are studying, the region
of parameter space which explains an eventual excess in $\sigma(pp\to A \to Zh
\to Zb\bar{b})$ is in the wrong-sign limit, the coupling of the lightest Higgs boson
to down-type quarks being of an opposite sign to the corresponding \sm{} quantity. Furthermore, the
relevant range of $\tan\beta$ values implies that, for the pseudoscalar,
the $b\bar{b}$ production mechanism is as important as the gluon-fusion
process. We will now show what such an excess would imply for the two other non-\sm{}-like
scalars of the theory -- $H$ and $H^\pm$.
\begin{figure}[t]
\begin{tabular}{cc}
\includegraphics[height=6cm,angle=0]{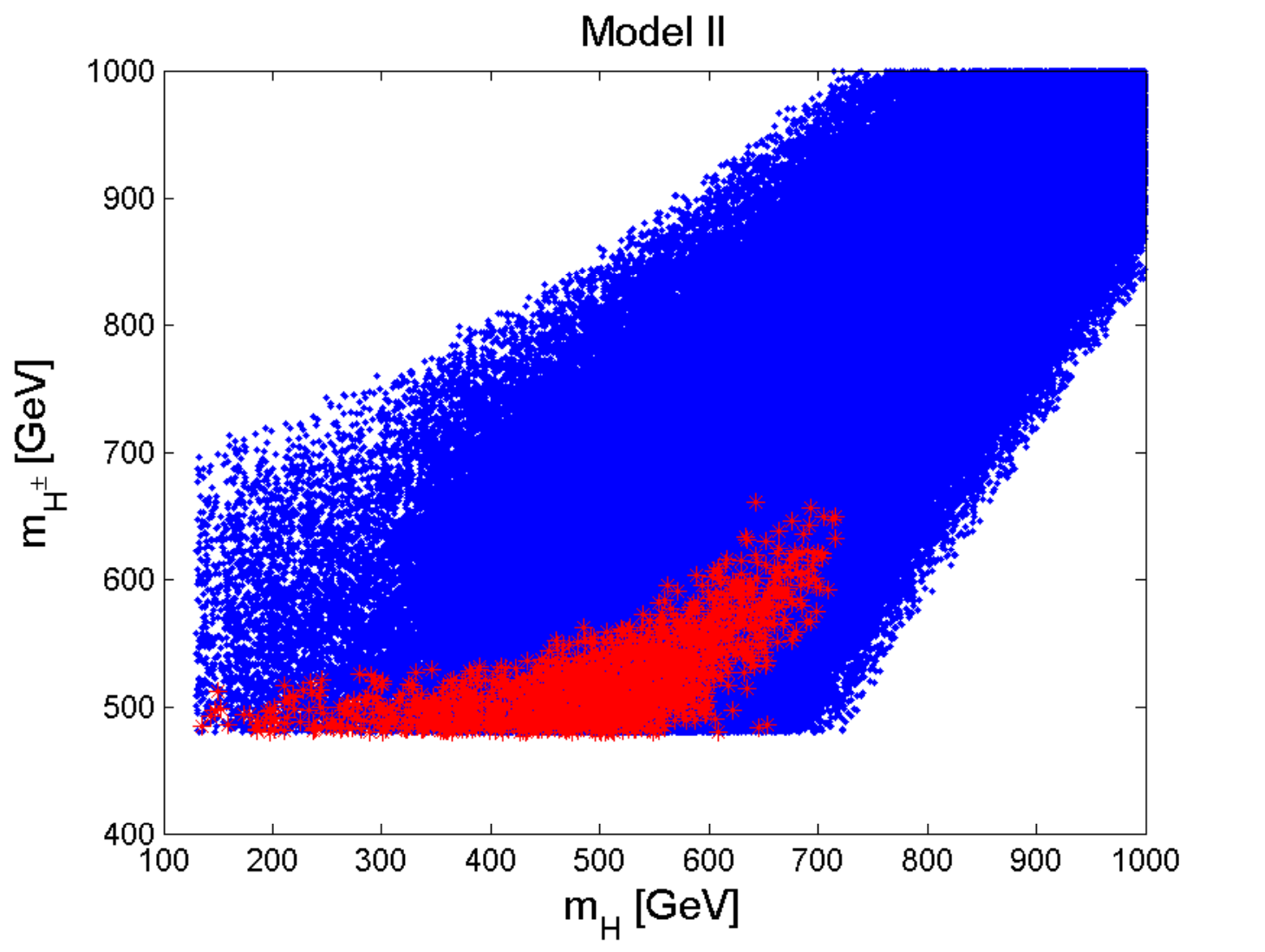}&
\includegraphics[height=6cm,angle=0]{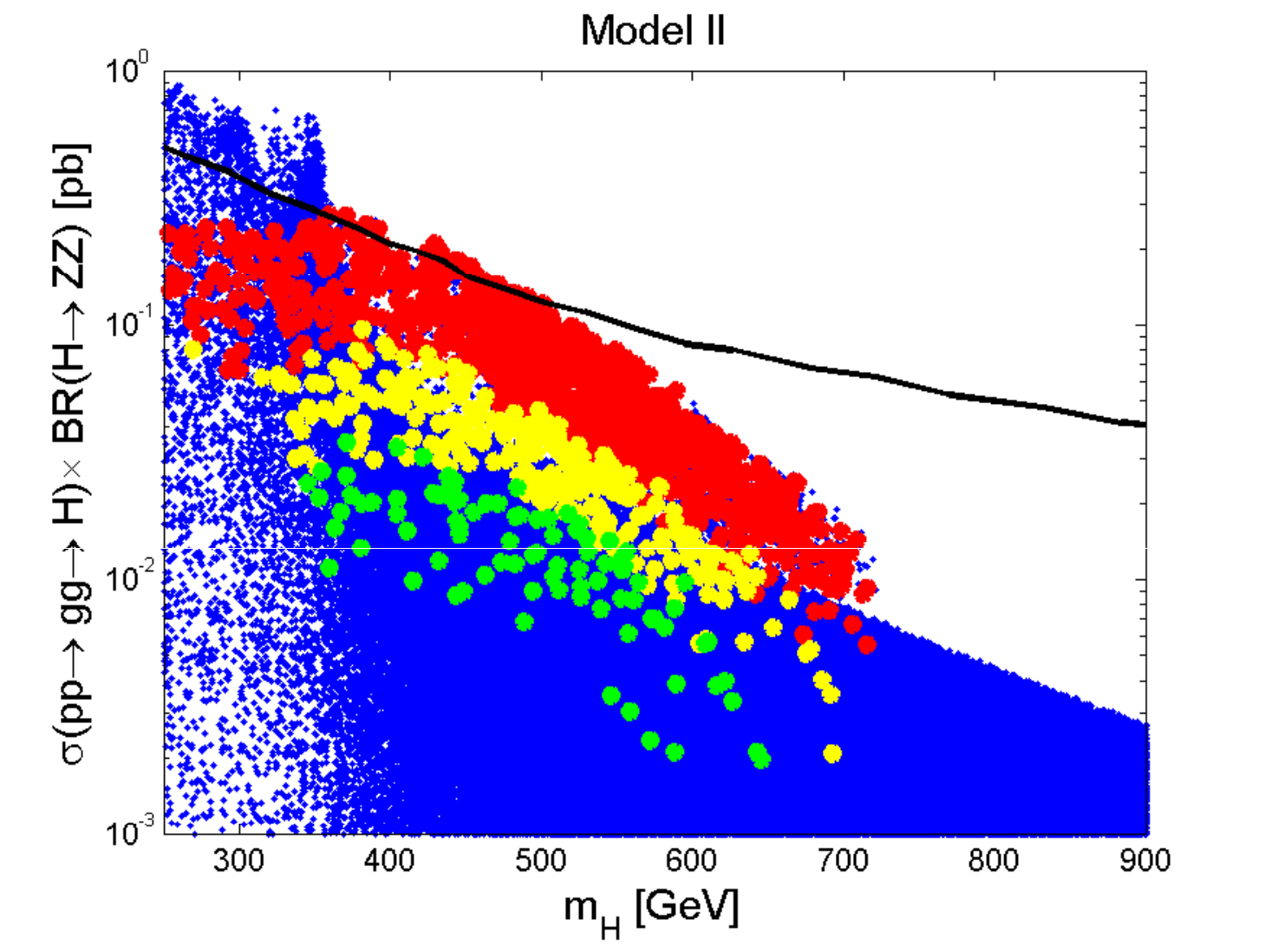}\\
 (a) & (b)
\end{tabular}
\caption{Scatter plots for the \thdm{} of type II showing
(a) the charged Higgs mass~$m_{H^\pm}$ versus the heavier \cp{}-even Higgs mass~$m_H$, both in GeV, and
(b) the gluon-fusion induced cross section for $gg\to H \to ZZ$ as a function of $m_H$ in GeV.
All relevant constraints are imposed and $h$ rates are within $20\%$ of their \sm{} values.
In (a) red points further satisfy $400\leq m_A\leq 500$ GeV and $\sigma(pp\to A \to Zh\to Zb\bar{b}) \geq 0.1$ pb.
In (b) red, yellow and green points are in the wrong-sign limit
and distinguish different regions of $\tan\beta$: $1<\tan\beta<5$ (red), $5<\tan\beta<7.5$ (yellow) and $\tan\beta>7.5$ (green).
The black line in (b) corresponds to the upper
$2-\sigma$ bound from the \atlas{} analysis~\cite{ATLAS-CONF-2017-058} based on an integrated
luminosity of $36$fb$^{-1}$.}
\label{fig:HpmHZZ}
\end{figure}

We first consider \fig{fig:HpmHZZ}~(a). We show the allowed values of
the charged scalar mass $m_{H^\pm}$ and the heavy \cp{}-even Higgs mass, $m_H$.
Clearly an excess in the channel $pp\to A \to Zh
\to Zb\bar{b}$ would require the two remaining scalars to have masses below about 700 GeV --
as was to be expected, since the wrong-sign limit cannot occur for very large
masses~\cite{Ferreira:2014naa}. The lower limit on $m_{H^\pm}$ observed in this figure
was imposed from the start, and pertains, as explained above, to bounds on $b\to s\gamma$.
In fact, the most recent results from $B$-physics calculations~\cite{Misiak:2017bgg}
already set this lower bound at $580$\,GeV -- which, we observe, could already be in
conflict with further constrained points, {\em i.e.} those for which the lightest
\cp{}-even scalar does not deviate from \sm{} behavior by more than $10\%$. Nonetheless,
if the $pp\to A \to Zh \to Zb\bar{b}$ excess were confirmed, this figure would give us the mass
range upon which the remaining \thdm{} scalars would have to be found, their masses at most
$\sim (150-300)$\,GeV apart.

As we argued previously
the coupling of the heavy \cp{}-even Higgs boson to two heavy gauge bosons is proportional
to $\cos(\beta-\alpha)$, and similarly the decay into two light \cp{}-even Higgs bosons
increases with $\cos(\beta-\alpha)$. It is thus not surprising that such
decays are very sensitive to the wrong-sign limit as well.
\fig{fig:HpmHZZ}~(b) displays the expected values for the \thdm{} cross section
for the gluon-induced process $gg\to H \to ZZ$. Such searches for heavy Higgs bosons
decaying into heavy gauge bosons are currently being probed at the
\lhc{} by both the \atlas{}~\cite{Aaboud:2017fgj,Aaboud:2017itg,Aaboud:2017gsl,ATLAS-CONF-2016-079,ATLAS-CONF-2017-058}
and \cms{}~\cite{CMS-PAS-HIG-16-001,CMS-PAS-HIG-16-033,CMS-PAS-HIG-16-034} collaborations.
The black line shown in this figure corresponds to to the upper $2$-$\sigma$ bound from nonobservation
of the recent \atlas{} analysis~\cite{ATLAS-CONF-2017-058} -- the \cms{} results would not be significantly
different -- and therefore gives us an idea where the current experimental
sensitivity to this process is. What is clear from the displayed figure is that the points
corresponding to the wrong-sign limit lie very closely to the black line. Therefore, the \thdm{} explanation for an
excess in $pp\to A \to Zh$ would imply a signal in the channel
$pp\to H \to ZZ$ of a magnitude that can surely be probed in the next years.
At larger values of $\tan\beta$ the constraints originating from the gluon-induced
channel are also weaker. It is worth
mentioning that in type II models the coupling of $H$ to bottom quarks is proportional to
$\cos\alpha/\cos\beta$. Since \sm{}-like behavior for the light Higgs boson~$h$ implies $\beta - \alpha \simeq \pi/2$,
this coupling is essentially growing as $\tan\beta$. As such, the $b\bar{b}$ production
process for $H$ will have a significant contribution, as it did for $A$ production,
and can thus constrain the regions with larger values of $\tan\beta$.

At $13$\,TeV both \atlas{} and \cms{} collected various constraints on
production cross sections of heavy Higgs bosons decaying into a pair of light Higgs
bosons at $125$\,GeV. Those searches include various di-Higgs final states:
$b\bar b\gamma\gamma$~\cite{ATLAS-CONF-2016-004,CMS-PAS-HIG-17-008},
$b\bar bb\bar b$~\cite{Aaboud:2016xco,CMS-PAS-B2G-16-026,ATLAS-CONF-2016-049}, $W^+W^- \gamma\gamma$~\cite{ATLAS-CONF-2016-071},
$b\bar b W^+W^-/ZZ$~\cite{Sirunyan:2017guj} and $b\bar b\tau^+\tau^-$~\cite{Sirunyan:2017djm}.
Still, these searches cannot yet test conclusively the wrong-sign limit:
for the parameter space region where the $pp\to A \to Zh$ excess occurs, the values for
the production cross section of $pp \to H \to hh$  are an order of magnitude below the current
experimental bounds, so we refrain from showing the corresponding figures.

\begin{figure}
\begin{center}
\includegraphics[height=6cm,angle=0]{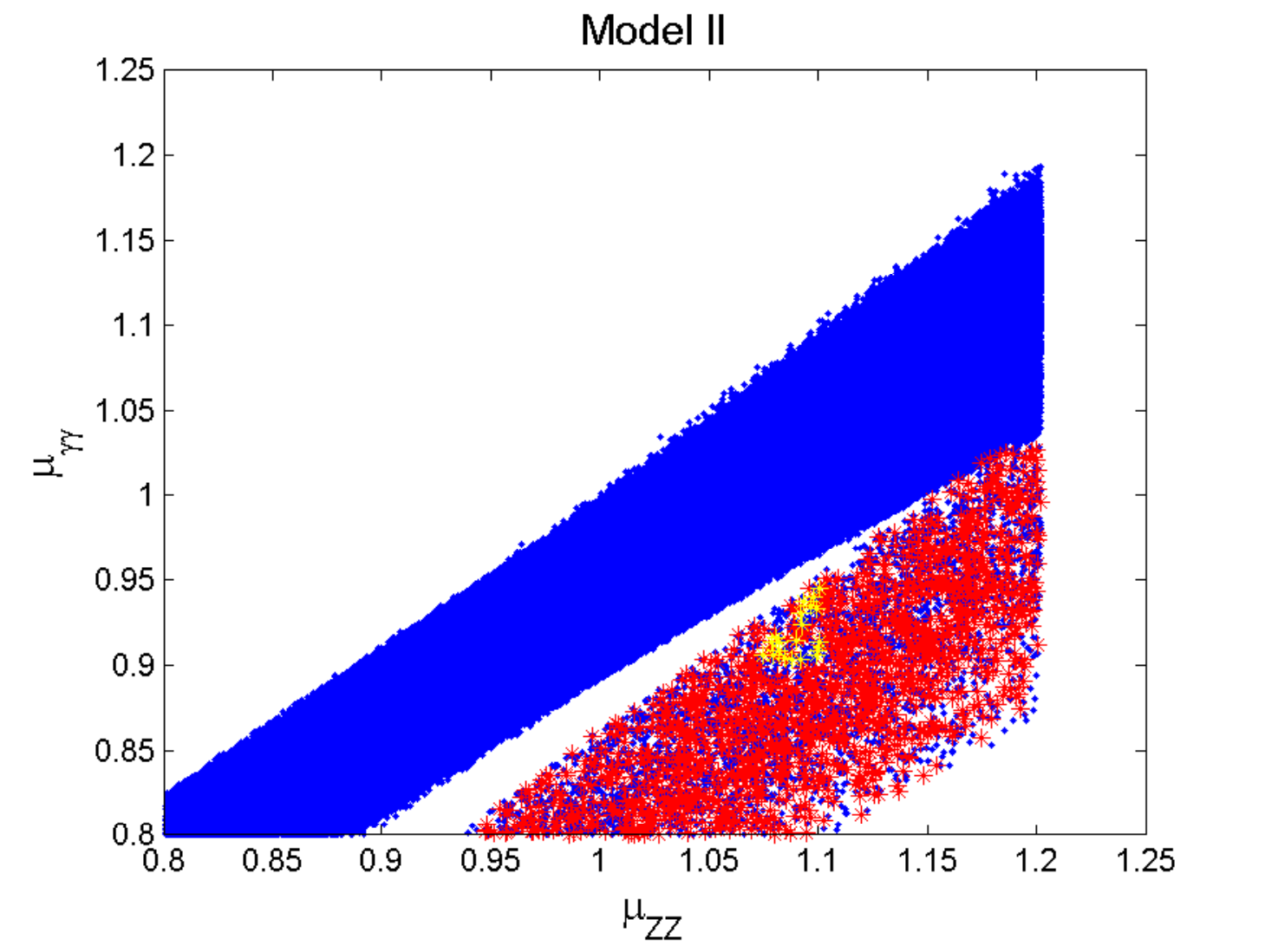}
\end{center}
\caption{Di-photon decay rate $\mu_{\gamma\gamma}$ versus
$ZZ$ production and decay rate $\mu_{ZZ}$ for the lightest \cp{}-even Higgs boson in the \thdm{} of type II.
It applies the same colour code as in \fig{fig:ppAZh}, highlighting in yellow
points where the rates of the lightest \cp{}-even Higgs-boson deviate from the \sm{} values by at most $10\%$.
}
\label{fig:ggZZ}
\end{figure}

Finally, let us remember that the wrong-sign limit is a nondecoupling regime. There are ``irreducible"
contributions to the gluon-fusion cross section and the di-photon decay width which make it
so that the light Higgs boson~$h$ can never have production and decay rates {\em exactly} like those of the \sm{} Higgs.
For the di-photon width, this stems from a destructive contribution to the amplitude from the charged scalar,
essentially independent of $m_{H^\pm}$ for masses up to $\sim 700$ GeV, see \citere{Ferreira:2014naa}.
As such the wrong-sign limit can be ruled out if measurements of sufficient precision are performed on the lightest
Higgs rates. To illustrate this, consider \fig{fig:ggZZ}, where we present the signal strength $\mu_{\gamma\gamma}$
versus $\mu_{ZZ}$ for the lightest Higgs boson.

The lower band of points in this figure corresponds to the wrong-sign limit, and it is clear it does not include
the point $(1,1)$ corresponding to the \sm{} expected values. Further, we see that the yellow points,
corresponding to an excess in $\sigma(pp\to A \to Zh \to Zb\bar{b})$ of at least $0.1$\,pb and \sm{}-like signal strengths within $10\%$, would
imply a di-photon rate for $h$ at least $5\%$ {\em smaller} than its \sm{} expectation, and a $ZZ$ rate at
least $7\%$ {\em larger} than the \sm{} value. Similar enhancements are predicted for the $WW$, $\tau\bar\tau$ and
$b\bar{b}$ channels, all of them arising from a positive interference in the wrong-sign limit between the top and bottom quark
contributions to the gluon-fusion cross section\footnote{Please notice that the amount of
enhancement of the gluon-fusion cross section in the wrong-sign limit is lowered by the inclusion
of higher-order \qcd{} corrections with respect to the \lo{} cross section~\cite{Ferreira:2014naa},
as observed for the reduction of the cross section in the \sm{} case.\label{foot:f}}.

\section{Conclusions}
\label{sec:conclusions}

We conclude that the wrong-sign limit in the \thdm{} of type II leads to enhanced and often dominant branching ratios of $A\to Zh$ and $H\to VV$, due to sizeable values of $\cos(\beta-\alpha)$.
It is thus possible to produce an excess in the current searches of heavy Higgs bosons decaying into Higgs and/or gauge bosons at the \lhc{}.
We demonstrated this enhancement for $pp\to Zb\bar{b}$, stemming from the production of a pseudoscalar $A$ with mass above the $t\bar{t}$ threshold and its subsequent decay to $Zh$, the \sm{}-like Higgs then decaying further to a pair of bottom quarks.
This statement applies to moderate values of $\tan\beta$ -- below $15$, above roughly $3$ -- which are experimentally not yet excluded by searches for heavy Higgs bosons decaying into fermions. At intermediate values of $\tan\beta$ between $\sim 5$ and $\sim 7.5$ the signal in the bottom-quark induced process is likely to be accompanied by a signal in the gluon-fusion process.
The wrong-sign limit makes testable predictions also for the lightest \cp{}-even Higgs boson couplings, namely enhancing the gluon fusion cross section and lowering the decay rate into two photons.
The mild excess currently seen by the \atlas{} collaboration in $pp\to Zb\bar b$ is entirely in agreement with the current precision achieved for the couplings of the lightest Higgs boson.

\section*{Acknowledgements}
This work began during the {\em Higgs Days at Santander -- 2017} workshop.
We thank Sven Heinemeyer for financial support and providing a rich discussion
environment and Hotel Chiqui for a renovated lobby with decent couches.
S.L. acknowledges support from ''BMBF Verbundforschung Teilchenphysik`` under grant number 05H15VKCCA.
J.W. acknowledges financial support by the PIER Helmholtz Graduate School.

{\footnotesize
\bibliographystyle{utphys}
\bibliography{AZh_2HDM}
}

\end{document}